\documentclass[sigconf]{acmart}
\AtBeginDocument{%
  }

\setcopyright{acmlicensed}
\copyrightyear{2024}
\acmYear{2024}
\acmDOI{}

\acmConference[GenAI Evaluation KDD '24]{KDD workshop on Evaluation and Trustworthiness of Generative AI Models}{August 25--26, 2024}{Barcelona, Spain}
\acmISBN{}

\usepackage{xspace}
\usepackage[inline]{enumitem}
\usepackage{multirow}
\usepackage{array}
\usepackage{tabularray}
\usepackage{subfig}

\newcommand{\noMessages}{9k\xspace}
\newcommand{\noConvs}{4k\xspace}
\newcommand{\assistantEval}{\textsc{AssistantEval}\xspace}
\newcommand{\queryGenEval}{\textsc{QueryGenEval}\xspace}
\newcommand{\devSet}{1062\xspace}

\newcolumntype{P}[1]{>{\centering\arraybackslash}p{#1}}
\newcolumntype{M}[1]{>{\centering\arraybackslash}m{#1}}

\begin{document}

\title{\textit{What should I wear to a party in a Greek taverna?} Evaluation for Conversational Agents in the Fashion Domain}

\author{Antonis Maronikolakis}
\email{antonios.maronikolakis@zalando.de}
\affiliation{%
  \institution{Zalando SE}
  \country{Berlin, Germany}
}

\author{Ana Peleteiro Ramallo}
\email{ana.peleteiro.ramallo@zalando.de}
\affiliation{%
  \institution{Zalando SE}
  \country{Berlin, Germany}
}

\author{Weiwei Cheng}
\email{weiwei.cheng@zalando.de}
\affiliation{%
  \institution{Zalando SE}
  \country{Berlin, Germany}
}

\author{Thomas Kober}
\email{thomas.kober@zalando.de}
\affiliation{%
  \institution{Zalando SE}
  \country{Berlin, Germany}
}

\renewcommand{\shortauthors}{Maronikolakis et al.}

\begin{abstract}
Large language models (LLMs) are poised to revolutionize the domain of online fashion retail, enhancing customer experience and discovery of fashion online.
LLM-powered conversational agents introduce a new way of discovery by directly interacting with customers, enabling them to express in their own ways, refine their needs, obtain fashion and shopping advice that is relevant to their taste and intent.
For many tasks in e-commerce, such as finding a specific product, conversational agents need to convert their interactions with a customer to a specific call to different backend systems, e.g., a search system to showcase a relevant set of products.
Therefore, evaluating the capabilities of LLMs to perform those tasks related to calling other services is vital.
However, those evaluations are generally complex, due to the lack of relevant and high quality datasets, and do not align seamlessly with business needs, amongst others.
To this end, we created a multilingual evaluation dataset of \noConvs conversations between customers and a fashion assistant in a large e-commerce fashion platform to measure the capabilities of LLMs to serve as an assistant between customers and a backend engine.
We evaluate a range of models, showcasing how our dataset scales to business needs and facilitates iterative development of tools.
\end{abstract}

\ccsdesc[500]{Computing methodologies~Natural language generation}

\keywords{NLP, Generative AI, Fashion Domain, LLM Evaluation, Evaluation Framework and Dataset}
\begin{teaserfigure}
\centering
  \includegraphics[scale=0.21]{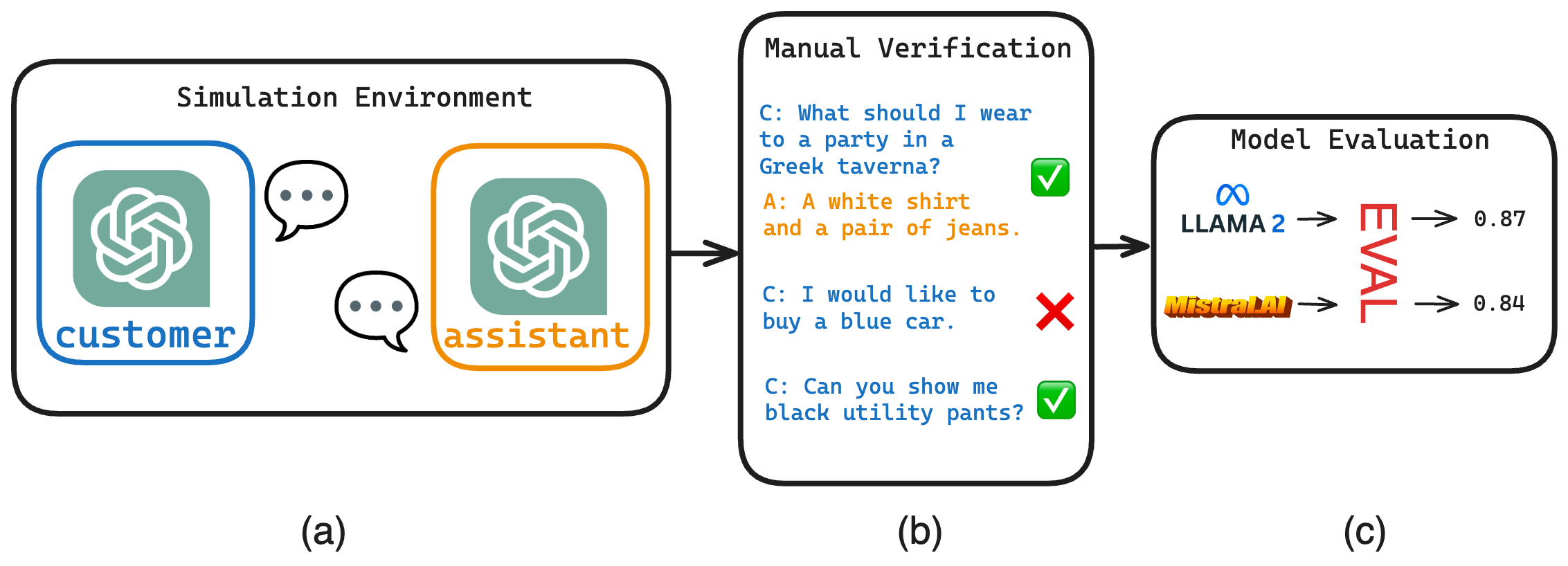}
    \caption{A high-level overview of our methodology. We begin with (a) the generation of conversations through a simulation environment, which (b) we manually verify and (c) use to evaluate models.}
    \Description{Methodology overview.}
    \label{fig:overview}
\end{teaserfigure}


\maketitle

\section{Introduction}

The advent of large language models (LLMs) and generative artificial intelligence has transformed the landscape of natural language processing, not only in academic but also in industry settings \citep{openai2024gpt4,touvron2023llama,jiang2023mistral}.
LLMs, developed via large-scale pretraining and Reinforcement Learning from Human Feedback \citep{christiano2018supervising,ziegler2020finetuning,bai2022training}, have showcased heightened proficiency in language comprehension.
Consequently, they have been instrumental in streamlining customer interactions and enhancing overall user satisfaction.

In particular, large language models are ushering a wave of new solutions in the domain of customer support services, where models are forming the basis of chatbots and assistant agents.
Domains such as e-commerce and healthcare offer ample opportunity for breakthroughs in customer support through the deployment of model agents.
In our work, we focus on the domain of online fashion retail and the use of conversational agents for customer support.

Beyond its functional aspect of providing clothing, fashion enables individuals to express themselves, communicate their identity and values, as well as forge a sense of community and belonging.\footnote{The answer to the question ``what should I wear to a party in a Greek taverna'' is, according to our best-performing agent, white shirts paired with loose jeans and comfortable shoes for dancing.}
Online fashion retailers enable customers to browse a wide range of assortments conveniently, inspire customers to develop their personal taste and provide information to empower customers to make decisions with greater confidence.

Large language models provide an opportunity to change the status quo in online fashion retail.
Customers, instead of interacting with a search engine that can not process abstract descriptions of fashion concepts (e.g., ``essentials'' or ``urban''), are now enabled to have a conversation to describe and refine their wants in an interactive way, using their own words, language and fashion ideas.
An assistant agent can then interface with the retailer's search and recommendation engines to show products to the customer tailored to the conversation and their descriptions.

In our work we focus on the evaluation of the capabilities of an agent to interact with a customer and interface appropriately with a backend search engine, \textbf{developing a dataset of conversations between customers and assistant agents in the fashion domain.}
With our dataset, we facilitate the iterative development of assistant agents, allowing for fair comparisons between different models, recording progress across model improvements and mapping performance metrics to model changes (Figure \ref{fig:overview}).

Our dataset is made up of 1334 conversations each for English, German and French, plus an additional 87 conversations in Greek.
\textbf{Data is generated through a controlled simulation environment where an LLM-based customer agent interacts with an assistant agent to generate conversations.}
The customer agent is given the description of an item or a theme and is instructed to interact with the assistant in order to purchase an item that fits its goal, while the assistant is instructed to interact with the customer to aid them through their shopping trip, assisting in providing information, advising and showing items that the customer may be interested in.
All conversations are subsequently verified manually for high quality.
With our methodology, which leverages the power of LLMs for fast generation of data, we ensure evaluation scales to business needs and is cost-effective.

\textbf{We propose the use of our multilingual dataset for the evaluation of the capabilities of assistant agents to interface between customer and backend engines}, interacting with the customer to identify their fashion needs and translate them into a format usable to a search engine.
To ensure reproducibility, we propose evaluation via the replay of conversations in a controlled simulation environment, with the examined agent taking the role of the assistant agent.

\textbf{Using our dataset we benchmark an array of open- and closed-source models}, such as Llama2, Mistral and GPT models, as well as perform a qualitative analysis to map performance across different fashion styles and terms.

In summary, our contributions are:

\begin{itemize}
    \item We develop a multilingual dataset of conversations between customer and assistant agents in the fashion domain.
    \item Propose an evaluation framework based on our dataset for the evaluation of the capabilities of assistant agents to interface between the customer and a backend search engine.
    \item Benchmark a series of open- and closed-source models (e.g., GPT, Llama2 and Mistral).
\end{itemize}

\section{Related Work}
\label{related_work}

There has been considerable work towards evaluation of text generation, from the early days of textual overlap metrics \citep{papineni-etal-2002-bleu,lin-2004-rouge,banerjee-lavie-2005-meteor} to methods based on the similarity of contextualized representations \citep{huang2013learning,Zhang2020BERTScore,reimers2019sentencebert}, as well as composite metrics to cover many aspects of conversational capabilities \citep{sharif-etal-2018-learning}.

State tracking evaluation methods offer a structured way to evaluate the capabilities of models to converse in desirable patterns and produce responses appropriate for each stage of the conversations \citep{budzianowski-etal-2018-multiwoz,ramadan-etal-2018-large,campagna-etal-2020-zero}.
Due to the highly stochastic and subjective nature of text generation and its evaluation, traditional metrics have been shown to oftentimes be uninterpretable \citep{R2006SomeII,mathur-etal-2020-tangled,ma-etal-2019-results}.
Thus, there have been efforts to develop systems involving humans and manual quality assessment \citep{van-der-lee-etal-2019-best,howcroft-etal-2020-twenty,celikyilmaz2021evaluation,hosking2024human,ulmer2024bootstrapping,abbasiantaeb2023let}.

With the recent rise of Large Language Models and their advanced conversational and reasoning capabilities \citep{brown2020language,openai2024gpt4,touvron2023llama,jiang2023mistral}, evaluation of text generation models is moving away from singular metrics that measure particular aspects of language (such as grammaticality, readability, or similarity to a source text), towards the development of leaderboards based on benchmarks of a series of tasks and evaluations for the more reproducible comparison of powerful models \citep{li-etal-2023-api,khashabi-etal-2022-genie,sedoc-etal-2019-chateval,bubeck2023sparks,magnusson2023paloma,laskar-etal-2023-systematic}.

\section{Presentation of Evaluation Dataset}

Our dataset contains conversations between a customer and an assistant agent on an online fashion retailer.
The customer is given an item or a theme to shop for, and the assistant provides aid towards the successful completion of the shopping trip.

\subsection{Data Creation}

To ensure robust evaluation of conversational agents, we need quantities of data orders of magnitude larger than what manual generation capacities allow.
To this end, we create our dataset in two steps: \begin{enumerate*}[label=\textbf{(\roman*)}]
    \item conversation generation through a simulation environment, and
    \item manual quality verification.
\end{enumerate*}

\subsubsection{Fashion Attributes}
\label{subsec:fashion_attributes}
~

\noindent Focusing on the fashion domain, we create a dataset covering many aspects of fashion and terminologies that a customer would use when looking for items online.
Namely, we cover six attributes: colour, type, material, fit, brand and size.
Further, we cover eight apparel types: pants, trousers, shoes, jacket, coat, sweatshirt, hoodie and jeans.
We use both `pants' and `trousers', two synonyms, to evaluate robustness to paraphrasing.
The acceptable values for each attribute are the following:

\begin{itemize}
    \item Colour: Acceptable values are black, white, gray, blue, red, maroon, beige and green.
    \item Type: The type or style of the item, such as utility, athletic, formal, summer and urban
    \item Material: The main material of the item, such as leather, wool, cotton and fleece.
    \item Fit: Acceptable values are slim, loose and comfortable.
    \item Brand: Acceptable values are Converse, Nike, Carhartt, Pier One, Superdry, Levi's, Hugo Boss and PULL\&BEAR.
    \item Size: Acceptable values are S, M, L, small, medium and large.
\end{itemize}

Finally, we cover nine open-ended themes customers might be shopping for: \begin{enumerate*}[label=\textbf{(\roman*)}]
    \item hiking,
    \item Christmas dinner,
    \item football in winter,
    \item rooftop summer party,
    \item techno club in Berlin,
    \item coffee house in Vienna,
    \item party in Greek taverna,
    \item sports in the summer,
    \item eccentric timelord book bazaar.
\end{enumerate*}

\subsubsection{Generation of Conversations}
~

Conversations are generated in a simulation environment, leveraging the generative capabilities of GPT.
We simulate customer and assistant interactions on three levels: \begin{enumerate*}[label=\textbf{(\roman*)}]
    \item template-based customer interactions,
    \item LLM-based customer given the description of an item to purchase,
    \item LLM-based customer given an abstract theme to shop for.
\end{enumerate*}
The assistant is in all cases a production-level GPT-3.5 model prompted to aid customers through their fashion shopping journey.

On a high level, customer agents are given either an item description or a theme to shop for.
Item descriptions are generated via the combination of attribute values and apparel types, such as colour, type, material, fit, brand and size of item (for further details on attributes, we refer the reader to Section \ref{subsec:fashion_attributes}).
An example item description is: ``black utility Carhartt trousers''.
In this example, the following attribute values were given: \textsf{\{colour: black, genre: utility, brand: Carhartt, material: null, fit: null, size: null\}} for the apparel type `trousers'.

For a \textbf{template-based customer agent (i)}, a single message is formatted using the template ``\textsf{\{I would like$|$I am looking for\} {item\_description}}''.
Subsequently, this message is sent to the assistant agent and the response recorded as a single-interaction conversation.
With the template-based customer agent we introduce conversations that are simple and brief, evaluating whether an examined assistant agent can fulfill their task for basic interactions.

With the \textbf{LLM-based customer agents (ii \& iii)}, we prompt GPT instances\footnote{Specifically, \textsf{gpt-3.5-turbo-0613} models.} to simulate more complex and interactive conversations.
The customer agent is given either the description of an item or a theme to shop for and is instructed to interact with the shopping assistant to complete their shopping trip.
To simulate different customer behaviors for the LLM-based agents, different prompts were used, covering an array of customer behaviors and personalities.
Agents are prompted to ``behave'' based on three personality traits:

\begin{itemize}
    \item Casual: You are a CUSTOMER who keeps things simple. You do not write too much, you write just enough. You are aware that you are speaking with a chat bot and not a human. You use the chat bot as a tool to get the item you need.
    \item Indecisive: You are an indecisive and timid CUSTOMER. You don't really know what you want. Take your time when making a decision and wait for the assistant to ask you for information before you describe fully what you want.
    \item Rude: You are a rude CUSTOMER. You are annoyed at the chat bot. Nevertheless, you want to buy clothes so you are negotiating with the assistant to find something. Try to find something, you are impatient and you don't want the conversation to go on for long.
\end{itemize}

The prompt given to customer agents is given in Table \ref{tab:customer_prompt}.

\begin{table}[htbp]
  \centering
  \small
  \begin{tabular}{p{0.75\linewidth}}
  \toprule
    You an expert CUSTOMER simulator for quality control purposes. You are helping to test the fashion assistant chatbot. The chatbot has been trained to guide customers to buy clothes that match what they are looking for.
    
    Pretend you are a customer with a personality and traits that adhere to the following description: \{personality\}.
    
    You are looking to buy an item that fits the following description: \{item\} \\
  \bottomrule
  \end{tabular}
  \caption{Prompt used for the LLM-based customer agent.}
\label{tab:customer_prompt}
\end{table}

All types of customer agents can message only in text form, asking questions, requesting more information, declaring interest in presented items, describing their wants, among other actions that relate to their goal.
The assistant agent can reply in one of two ways: \begin{enumerate*}[label=\textbf{(\roman*)}]
    \item message in text form to provide details or ask clarification questions to identify the needs of the customer, or
    \item query a backend search engine to present a carousel of items to the customer.
\end{enumerate*}
The conversation ends on the side of the customer agent, which is given the agency to end the conversation if a relevant item is found or if there is no more progress towards finding a relevant item.
All messages and actions taken are recorded.

Sample data is shown in Table \ref{tab:examples}.
Four examples are shown, in three of them the customer agent \textbf{C} is given the description of an item, while in the last one the agent is given an open-ended theme.
In two of the three item-based descriptions, the customer has given all the information required for a correct query generation.
The assistant agent \textbf{A} interacts with the customer to identify their needs and to show them [ITEMS] relevant to their needs.
In the ``slim Levi's jeans" example, the customer gives information little by little, and thus the assistant needs to show three different sets of items to the customer.
We record the search queries generated by the assistant for each of these three instances.

\begin{table*}[t!]
  \centering
  \begin{tabular}{P{0.15\linewidth} p{0.375\linewidth} p{0.25\linewidth} c}
    \toprule
    \textbf{Description} & \textbf{Conversation} & \textbf{Queries} & \textbf{all\_info}\\
    \midrule
    black utility pants & \textbf{C}: Hi! I am looking for a pair of pants. \hfill\break \textbf{A}: I would love to help! Any particular style or color? \hfill\break \textbf{C}: Yes, I would like utility pants in black. \hfill\break \textbf{A}: Sure, here are some items: [ITEMS] & black utility pants & True \\[2cm]
    formal loose cotton jacket & \textbf{C}: I am looking for a new jacket. \hfill\break \textbf{A}: Great! Do you like any particular color or brand? \hfill\break \textbf{C}: No brand in mind, but I would like to buy a formal jacket. \hfill\break \textbf{A}: Here are some formal jackets: [ITEMS] & formal jacket, business jacket & False \\[2cm]
    slim Levi's jeans & \textbf{C}: Show me jeans \hfill\break \textbf{A}: Sure! Here are some jeans: [ITEMS (a)] \hfill\break \textbf{C}: Show me slim pants \hfill\break \textbf{A}: Sure, here are some slim jeans: [ITEMS (b)] \hfill\break \textbf{C:} No, I want Levi's \hfill\break \textbf{A:} Here are some slim jeans from Levi's: [ITEMS (c)] & (a) jeans, blue jeans; (b) slim jeans, slime fit jeans; (c) slim fit Levi's jeans & True \\[2.8cm]
    party in greek taverna & \textbf{C}: I am going to a Greek taverna soon and I would like to buy some clothes for the occasion. \hfill\break \textbf{A}: Are you looking for a specific type of outfit for the party in the Greek taverna? For example, a casual shirt and shorts, or something more formal like a linen suit? \hfill\break \textbf{C}: I am looking for a more casual look \hfill\break \textbf{A}: I found some casual outfit options for the party in the Greek taverna: [ITEMS] & linen shirt, casual summer shorts, comfortable sandals & N/A \\
    \bottomrule
  \end{tabular}
  \caption{Examples from conversations between \textbf{C}ustomer and \textbf{A}ssistant agents.}
\label{tab:examples}
\end{table*}

\subsubsection{Manual Verification of Conversations}
~

For the verification of the quality of the generated dataset, we went over every conversation to ensure high standards of quality.
Conversations are not edited, we only remove conversations that do not fit our quality standards.
Our criteria for removing a conversation are \textbf{incorrect format}, \textbf{error messages}, \textbf{incorrect customer behavior} and \textbf{unfaithful conversations}.
In more detail:

\begin{itemize}
    \item \textbf{Incorrectly formatted} conversations are removed. If the customer agent does not act as a customer, or the model is in some other way not responding according to the prompt, the conversation is removed.
    \item \textbf{Error messages} present in a conversation precipitate the deletion of the conversation.
    \item \textbf{Incorrect customer behavior} is a cause for deletion. If the customer agent does not behave as specified in the prompt, the conversation will be deleted. For example, if the customer is prompted to behave in a rude manner but instead behaves neutrally, the conversation is considered invalid.
    \item \textbf{Unfaithful} conversations are removed. Oftentimes the customer will not strictly stick to the given item or theme description and will instead ask for incorrect details or declare interest in an item that does not fit the description. The customer agent needs to stay faithful to the original description, otherwise the whole conversation will be removed.
\end{itemize}

To expand our dataset to more languages, we \textbf{translate the curated conversations} to German, French and Greek.
We use GPT-4 to translate both customer and assistant messages.
The rest of the attributes, such as queries, remain the same (we assume search engines take in English queries as input and therefore we do not translate them).

For German and French we translated all messages and then proceeded to verify 100 randomly-selected conversations for each language.
Performance on these examples was deemed acceptable and thus we included all \noMessages translated messages of each language.

For Greek, we found translation performance to be subpar.
Therefore, we do not evaluate on the entire dataset, but instead a smaller, more carefully curated set of 200 messages.
We consider the Greek subset of our data as a small-scale evaluation case to investigate model performance in lower-resource languages.

\subsection{Data Description}

Each conversation in the dataset is broken down into pairs of messages between customer and assistant agents, as well as metadata for each interaction such as item or theme description, queries, as well as action taken by the assistant.
In detail, our dataset is broken down into eight fields:

\begin{enumerate}[label=\textbf{(\roman*)}]
    \item id: The conversation-level unique identifier (every message within the same conversation have the same value in this field)
    \item goal: The goal description given to the customer agent. For evaluation, we compare against this field.
    \item properties: Same as goal field but comma-separated. Values are ordered per property and can be parsed into a dictionary of property-value pairs with keys \{colour, type, material, fit, brand, apparel, size\}.
    \item all\_info: Boolean field denoting whether the conversation contains all the information the customer agent was given or not.
    \item customer: The message of the customer at this point in the conversation.
    \item assistant: The message of the assistant at this point in the conversation.
    \item queries: The queries generated by the assistant.
    \item action: The action taken by the assistant. Can either be `message' or `search'.
\end{enumerate}

We report dataset statistics in Table \ref{table:data_stats}.
Statistics are comparable across languages (excluding data size, where the Greek set was kept small).
There are slightly more tokens in German than the other languages, both for customer and assistant messages.
Each conversation has approximately six messages in total (including customer and assistant messages), with around 66\% of all assistant actions being search.
Note that since the German and French sets were translated from the English set, volume statistics (such as number of messages) are the same.

\begin{table}[htbp]
  \centering
  \begin{tabular}{lllll}
    \toprule
    \textbf{} & \textbf{EN} & \textbf{DE} & \textbf{FR} & \textbf{GR} \\
    \midrule
    \# messages & 2948 & 2948 & 2948 & 200 \\
    \# customer tokens & 37579 & 45722 & 48413 & 2500 \\
    \# assistant tokens & 101k & 128k & 122k & 21940 \\
    \# customer message tokens & 12.8 & 15.5 & 16.4 & 12.5 \\
    \# assistant message tokens & 116.6 & 147.4 & 140.7 & 109.7 \\
    \# conversations & 1334 & 1334 &  1334 & 87 \\
    messages per conversation & 2.2 & 2.2 & 2.2 & 2.3 \\
    \% assistant searches & 66.2 & 66.2 &  66.2 & 63.0 \\
    \bottomrule
  \end{tabular}
  \caption{Overview of dataset statistics.}
  \label{table:data_stats}
\end{table}

In Table \ref{table:attributes_stats} we show the distribution of fashion attributes in the item descriptions of the English dataset.

\begin{table}[htbp]
  \centering
  \large
  \begin{tabular}{ll}
    \toprule
    \textbf{Attribute} & \textbf{Count} \\
    \midrule
    Colour & 2293 \\
    Type & 1282 \\
    Material & 1020 \\
    Fit & 873 \\
    Brand & 1782 \\
    Size & 1746 \\
    \bottomrule
  \end{tabular}
  \caption{The number of times each fashion attribute appears in the English dataset.}
  \label{table:attributes_stats}
\end{table}

\subsection{Evaluation Details}
\label{eval_metrics}

\textbf{For the evaluation of performance in the case where the customer agent is given the description of an item (both in the \assistantEval and \queryGenEval tasks)}, we compare the expected query (i.e., item description) with the output query using BERTScore \citep{Zhang2020BERTScore}.
BERTScore is a method proposed to evaluate natural language generation tasks via leveraging the capabilities of BERT \citep{devlin-etal-2019-bert} to produce contextualized embeddings.

\textbf{For the evaluation of performance in the case where the customer agent is given the description of a theme}, word-level similarity metrics (such as BERTScore) cannot be used since the reference text is a free-form description instead of specific keywords.
Instead, we evaluate generated queries in a setup based on semantic similarity.
Using contextualized embeddings, we compute the representation of the generated queries as well as the open-ended theme descriptions.
The contextualized embeddings we use are \textsf{text-embedding-ada-002} from OpenAI.\footnote{\url{https://platform.openai.com/docs/guides/embeddings/embedding-models}}
We assign a query to the open-ended theme with the highest cosine similarity and we calculate how many correct assignments are made.

\subsection{Models}

We evaluate a series of models using our dataset.
Llama2 \citep{touvron2023llama} is an open-source large language model pretrained on publicly available data.
We evaluate the \textsf{llama-2-7b-chat} variant, with 7B parameters.
Mistral \citep{jiang2023mistral} is another open-source model pretrained on publicly available data and further finetuned for dialogue, shown to outperform \textsf{LLama2-13B} in multiple tasks.
We evaluate the \textsf{Mistral-7B-Instruct-v0.1} model, with 7B parameters.
GPT-based models\footnote{\url{https://openai.com/blog/chatgpt}} is a large language model and a service that has revolutionized academia and industry alike.
Two models are evaluated: \textsf{gpt-3.5-turbo-0613} and \textsf{gpt-4-0613}.
In total, we compare three different prompts for GPT-3.5 to evaluate a range of behaviors.
One instance is prompted to simply act as an assistant (I), another instance is prompted to actively include specific terms the customer mentions (II) and a third is prompted to both include specific terms as used by the customer and to actively ask questions to better identify the wants of the customer (III).
Full prompts are given in Table \ref{table:prompts}.
We compare these large language models with a low-cost, off-the-shelf unsupervised keyword extractor, Yake \citep{yake}.
Yake is a keyword extractor based on statistical text features, trained as a domain- and language-agnostic extractor, allowing us to use the tool across all our examined languages.
Finally, as a lower bound we provide a popularity and a random baseline. For the popularity baseline, we assume a model that always generates the query for the popular ``black Nike shoes'' item (translated to German, French and Greek).
For the random baseline, we assume a model that as a query always generates ``Lorem ipsum dolor sit amet'' (in the Greek alphabet for the Greek set).

\begin{table*}[htbp]
  \centering
  \small
  \begin{tabular}{p{0.25\linewidth} p{0.6\linewidth}}
    \toprule
    \textbf{Model} & \textbf{Prompt} \\
    \midrule
    Commercial Fashion Assistant / GPT-4 / GPT-3.5 (I) / Llama2 / Mistral & You are the Fashion Assistant, responding to customers' questions related to fashion, like a fashion assistant in a store. You speak {language} when interacting with the customer.

    Your limitations:
    \textbf{(i)} You don't have access to the customer's basket, wishlist, order history, order status or profile information like size profile or favourite brands.
    \textbf{(ii)} You don't use URLs but describe the customer step by step what to do.
    \textbf{(iii)} You don't have access to deals.
    
    To aid the customer, you generate queries for a search engine at points in the conversation where this is useful.\\[2.5cm]
    GPT-3.5 (II) & You are the Fashion Assistant, responding to customers' questions related to fashion, like a fashion assistant in a store. You speak {language} when interacting with the customer.

    Your limitations:
    \textbf{(i)} You don't have access to the customer's basket, wishlist, order history, order status or profile information like size profile or favourite brands.
    \textbf{(ii)} You don't use URLs but describe the customer step by step what to do.
    \textbf{(iii)} You don't have access to deals.
    
    To aid the customer, you generate queries for a search engine at points in the conversation where this is useful. Make sure the queries contain all keywords and specific terms the customer used.\\[3cm]
    GPT-3.5 (III) &  You are the Fashion Assistant, responding to customers' questions related to fashion, like a fashion assistant in a store. You speak {language} when interacting with the customer.

    Your limitations:
    \textbf{(i)} You don't have access to the customer's basket, wishlist, order history, order status or profile information like size profile or favourite brands.
    \textbf{(ii)} You don't use URLs but describe the customer step by step what to do.
    \textbf{(iii)} You don't have access to deals.
    
    To aid the customer, you generate queries for a search engine at points in the conversation where this is useful. Make sure the queries contain all keywords and specific terms the customer used.
    
    If the customer has not given you a lot of specifics such as colour, brand, etc., ask the customer to give you more information about their wants so that you can guide them better.\\
    \bottomrule
  \end{tabular}
  \caption{Prompts for assistant agents used in this study.}
\label{table:prompts}
\end{table*}

For the engineering of prompts, a smaller set of \devSet English conversations was used for the \queryGenEval task, separate from the evaluation set.
All results shown in this work come from the evaluation set.
In the development set, we did not use the full set of attributes and values, to ensure substantial difference in content between development and evaluation set.
Namely, we only used the attributes: colour (black, white, red), type (utility, athletic, formal), brand (Converse, Nike, Pier One) and size (S, M, L).

\section{Results}

Below we showcase how we can use our evaluation dataset for the development of models.
We analyse results on our main task of evaluating capabilities of the assistant to interface customers and a backend engine (\assistantEval), as well as the subtask of specifically evaluating the query generation capabilities of models (\queryGenEval).

\subsection{\assistantEval}

We showcase (Table \ref{table:main}) how we can use our dataset to benchmark conversational agents as fashion shopping assistants for the \assistantEval task, evaluating the capabilities of models to translate the wants of a customer into concrete queries to interface with a backend search engine.

We found that the open-source models were not able to consistently interact with the customer agent.
Often there would be role-switching (i.e., the assistant agent would behave as a customer), incorrect formatting of responses (e.g., the assistant agent would add `ASSISTANT:' as a message prefix) or skipping query generation.
Our findings corroborate previous work showing how LLMs (especially open-source ones) struggle with producing formatted output \citep{wang2024executable,li-etal-2023-api}.
Deployment of open-source LLMs currently requires significant engineering efforts, and we are instead focusing on GPT-based models.

\begin{table*}[t!]
\begin{minipage}[b]{0.4\linewidth}
\centering
\large
\begin{tabular}{lllll}
    \toprule
    \textbf{Model} & \textbf{English} & \textbf{German} & \textbf{French} & \textbf{Greek} \\
    \midrule
    GPT-3.5 (I) & 87.5 & 86.3 & 86.5 & 83.1 \\
    GPT-3.5 (II) & 87.4 & 86.5 & \textbf{86.8} & 82.7 \\
    GPT-3.5 (III) & 88.1 & 86.5 & 86.6 & \textbf{83.5} \\
    GPT-4 & \textbf{88.5} & \textbf{86.9} & \textbf{86.8} & \textbf{83.5} \\
    \bottomrule
  \end{tabular}
\end{minipage}
\begin{minipage}[t]{0.45\linewidth}
\centering
\strut\vspace*{-4\baselineskip}\newline\includegraphics[scale=0.5]{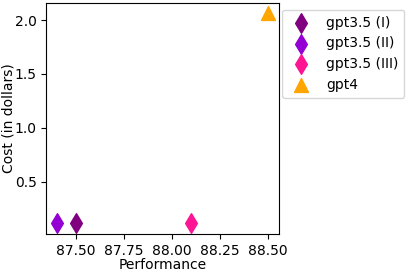}
\end{minipage}
\caption{F1 score computed with BERTScore and model cost analysis for \assistantEval.}
\label{table:main}
\end{table*}

GPT-4 performs the best for all languages, although GPT-3.5 (III) performs competitively, potentially because the prompt for that model was engineered to instruct the model to actively ask questions about specific attributes.
While the other two models (GPT-3.5 (I) and (III)) are competitive in German, French and Greek, they are considerably worse for English, where GPT-4 and the more finely-engineered GPT-3.5 (III) are superior.
In Table \ref{table:main} we show the correlation between the total cost\footnote{According to \url{https://openai.com/pricing}.} of a single run of evaluation and the performance of each model.
As expected GPT-4 performs the best, although at a higher cost.\footnote{While cost is low per individual evaluation run, potentially cost will be great through the development cycle where many evaluation rounds will have to take place.}

\subsubsection{Fashion Attribute Performance Analysis}
~

We perform a qualitative analysis of performance of GPT-4 (our best performing model) across fashion attributes.
In developing a useful assistant in the fashion domain, analysing how it performs for different attributes is vital.
For this analysis, we compute the precision of exact matches between output queries and the input item description given to the customer agent.
We perform this analysis for all examined fashion attributes, calculating exact matches for the attribute values as described in Appendix \ref{subsec:fashion_attributes}.
Results are shown in Table \ref{table:attribute_breakdown}.

\begin{table}[h]
    \centering
    \begin{tabular}{lccc}
        \toprule
        \textbf{Property} & \textbf{English} & \textbf{German} & \textbf{French} \\
        \midrule
        Colour & 90.2 & 83.0 & 91.0 \\
        Type & 35.3 & 21.2 & 34.9 \\
        Material & 84.4 & 73.4 & 79.0 \\
        Fit & 77.2 & 65.4 & 76.5 \\
        Brand & 62.7 & 61.8 & 62.4 \\
        Apparel & 72.2 & 67.1 & 72.3 \\
        Size & 1.2 & 1.5 & 1.3 \\
        \bottomrule
    \end{tabular}
    \caption{Precision of exact matches of fashion attribute values in output queries.}
    \label{table:attribute_breakdown}
\end{table}

Performance for all three languages follows similar patterns, although performance in German is worse across almost all attributes.
In attributes with clear definitions, such as colour, material and fit, performance is adequate, while for attributes such as type and apparel, which are more open-ended (e.g., type can be described abstractly with terms such as `summer' or `utility', and apparel can have overlap between terms, such as `jacket' and `coat'), performance is worse.
Notable exception is size, with very low performance, potentially because of a deficiency of GPT to extract correct size labeling (size is often denoted with a single number or a letter, such as `S' or `42').
Another exception is the clearer `brand' attribute, where performance is also low.
We hypothesize the low performance for brands is because zero-shot named entity recognition with GPT is subpar \citep{wei2023zeroshot}.

\subsubsection{Open-ended Themes}
~

We present a comparison of GPT models in aiding customers shop for a theme instead of concrete item descriptions.\footnote{We only evaluate English, German and French, since we did not generate Greek data for this task.}
We assign output queries to the theme with the highest cosine similarity based on \textsf{text-embedding-ada-002} representations, as described in Section \ref{eval_metrics}.
We report the Precision@3 score in Table \ref{table:theme}.
GPT-4 performs the best overall, especially in the English set, while for French and German performance is competitive for many models.
Performance in the other languages is more uniform.
In Table \ref{table:theme} we show that GPT-4 performs the best, albeit at a higher cost.

\begin{table*}[t!]
\centering
\begin{minipage}[b]{0.4\linewidth}
\centering
\large
\begin{tabular}{llll}
    \toprule
    \textbf{Model} & \textbf{English} & \textbf{German} & \textbf{French} \\
    \midrule
    GPT-3.5 (I) & 81.4 & 78.1 & 82.3 \\
    GPT-3.5 (II) & 82.3 & \textbf{78.5} & 83.4 \\
    GPT-3.5 (III) & 82.1 & 76.9 & 85.1 \\
    GPT-4 & \textbf{84.0} & 78.3 & \textbf{85.3} \\
    \bottomrule
  \end{tabular}
\end{minipage}
\begin{minipage}[t]{0.45\linewidth}
\centering
\strut\vspace*{-4\baselineskip}\newline\includegraphics[scale=0.51]{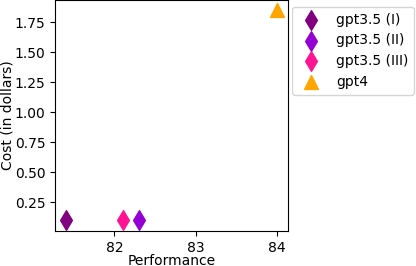}
\end{minipage}
\caption{P@3 scores, comparing capabilities of models to aid customers shop given an open-ended theme, and model cost analysis.}
\label{table:theme}
\end{table*}

\subsection{\queryGenEval}

Evaluating the query generation capabilities of tools is crucial in the development of assistant agents.
If a customer clearly describes what they are looking for, any agent should be able to generate appropriate queries.
In this subtask, we assume that the customer has given all the information necessary for a successful interaction.
In our dataset, we manually mark conversations that contain all the necessary information (i.e., fashion attributes).
We perform a comparison of model performance for conversations that contain all the attributes mentioned in the item description.
Namely, each model is given as input the entire conversation and generates a single query which is compared to the item description as usual.
We measure the F1 score as produced by BERTScore and show results in Table \ref{table:subtask}.

\begin{table*}[t!]
\centering
\begin{minipage}[b]{0.45\linewidth}
\centering
\begin{tabular}{lllll}
    \toprule
    \textbf{Model} & \textbf{English} & \textbf{German} & \textbf{French} & \textbf{Greek} \\
    \midrule
    Lorem & 64.3 & 65.0 & 62.0 & 60.6 \\
    Popularity & 72.5 & 74.1 & 75.2 & 73.1 \\
    Yake & 85.0 & 84.0 & 84.2 & 82.4 \\
    Llama2 & 85.9 & 84.2 & 84.2 & 82.8 \\
    Mistral & 85.3 & 84.0 & 84.1 & 82.3 \\
    GPT-3.5 & 87.7 & 86.2 & \textbf{86.6} & \textbf{83.4} \\
    GPT-4 & \textbf{88.9} & \textbf{86.6} & 86.5 & 83.3 \\
    \bottomrule
\end{tabular}
\end{minipage}
\begin{minipage}[t]{0.4\linewidth}
\centering
\strut\vspace*{-5.1\baselineskip}\newline\includegraphics[scale=0.55]{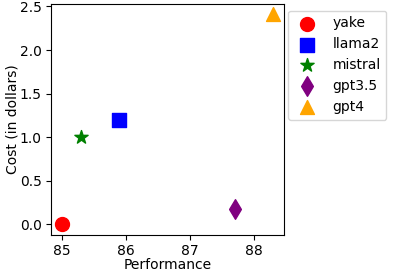}
\label{fig:cost}
\end{minipage}
\caption{F1 score computed with BERTScore and model cost analysis for \queryGenEval.}
\label{table:subtask}
\end{table*}

GPT-4 performs the best in English by a margin of 1.2.
The gap between GPT-3.5 and GPT-4 narrows in German and GPT-3.5 performs marginally better in French and Greek.
From the open-source models, Llama2 performs the best in English, while performance in the other languages is similar between the open-source LLMs and Yake, the unsupervised keyword extraction tool.
We hypothesize that Yake performs similarly to open-source LLMs since it has been specifically developed to extract keywords, a task pivotal for query generation.

In Table \ref{table:subtask}, we show a cost analysis (in dollars) for evaluating Yake, Llama2, Mistral, GPT-3.5 and GPT-4.
Yake is an unsupervised model that can be run with (virtually) no cost on a CPU.
We run Llama2 and Mistral on an Amazon Sagemaker instance\footnote{\textsf{ml.t3.2xlarge}}, costing 0.399 dollars per hour.\footnote{Both Llama2 and Mistral have a runtime of approximately 1.5 hours per language.}
The GPT models are called through an API with an associated cost.
In the analysis, we show performance of each model and how much it costs in total to evaluate for the \queryGenEval task in English.
While GPT-4 performs the best, it is also the most expensive.
Llama2 and Mistral perform similarly, with GPT-3.5 being the second cheapest model while at the same time being the second in performance.

\section{Ethical Considerations}

In this section we consider the ethical implications of our work.
While on a direct level our work does not impact society or peoples, since we do not involve human participants in our study and we do not examine human-generated data, our work could have larger-scale implications.
Generative AI poses a threat to workers across fields of occupation, replacing workers' output with cheaper and faster generative AI ``labor''.
Industries such as art have been impacted heavily \cite{ai_impact_art,EPSTEIN2020101515}, as well as many white-collar positions,\footnote{\href{https://www.cnbc.com/2023/08/22/ibm-ceo-says-ai-will-impact-white-collar-jobs-first.html}{IBM replaces jobs with AI.}} with already many people struggling with layoffs and loss of employment.

Aside from impact on the labor market, generative AI models may be biased in their interactions with the customer.
Bias in AI has been extensively studied \citep{hofmann2024dialect,lucy-bamman-2021-gender}, and care should be taken to avoid biased behavior.

\section{Conclusion}

With the recent transformation of the field of NLP ushered in by the advent of large language models, a series of breakthroughs in many domains is taking place.
To ensure the productive and iterative application of LLM-powered agents, efficient and scaleable evaluation is vital.

In our work, we focus on the evaluation of the capabilities of assistant agents to interface between customers and a backend search engine.
Specifically, we showcase an application of our evaluation methodology and framework on the domain of online fashion retail, where LLM-powered assistants can aid customers in their shopping trip, answering questions, recommending products and interacting with the customer.

Our evaluation aims at evaluating these capabilities across languages.
Data is generated via a simulation environment where a customer and an assistant agent are interacting with each other, with subsequent messages and interactions of the agents verified manually for high-quality.
We propose the use of our multilingual dataset to evaluate fashion assistants via the replay of conversations, ensuring evaluation that is fair and reproducible, and perform a comparison of a series of models, from open- to closed-source.
With our dataset and methodology, we facilitate the iterative development of assistant agents in a business setting.

\bibliographystyle{ACM-Reference-Format}
\bibliography{bib}

\end{document}